\documentstyle[11pt,newpasp,twoside,epsf]{article}

\begin{document}

\title{A 3D morpho-kinematic study of NGC 3132}

\author{Monteiro, H.$^1$, Morisset, C.$^2$, Gruenwald, R.$^1$, \&
Viegas, S.M.$^1$} 
\affil{$^1$Instituto Astron\^omico e Geof\'{\i}sico, USP,
Avenida Miguel Stefano, 4200 CEP 04301-904 S\~ao Paulo, SP, Brazil\\ 
$^2$Laboratoire d'Astronomie Spatiale, Traverse du Siphon, Les 
Trois Lucs, 13012 Marseille, France}

\begin{abstract}
We present a new modeling tool for planetary nebulae (PNe), 
based on 3D photoionization calculations. 
From models for two theoretical PNe, we show that the enhancement
in the equatorial zone observed in several PNe is not necessarily due
to a density gradient, as usually interpreted.
We study the morpho-kinematic properties of the PN
NGC~3132 and show that a bipolar Diabolo shape successfully
reproduces the observed images, as well as the high resolution observations. 
 
\end{abstract}
\vspace{-0.8cm}

\section{Introduction}
Planetary Nebulae show different morphologies, but since the
works of Kwok et al. (1978) and Balick (1987), 
it is possible to reproduce their observed shape basically with two
types of geometry: spherico-elliptical ones and bi-polar ones
(also called butterfly). 
The correlation between morphological types and
other properties linked to the central star mass set constraints on
stellar evolution theories (e.g. Stanghellini et al. 1993,
Corradi \& Schwarz 1995).
Here a new modeling tool for studying the morphology and kinematics of
PNe is presented. This tool is first used to show that an equatorial
brightness enhancement does not always correspond to a density
enhancement. Secondly we present two models for NGC~3132. The first one
is an ellipsoidal shell, as suggested in the literature, but which neither reproduces the density variation
along the nebula, nor the velocity profiles. A Diabolo shape is
successful in reproducing consistently all the observations. 

\vspace{-0.4cm}
\section{A new modeling tool}
The results of a 3D photoionization code 
(Gruenwald, Viegas \& Brogui\`ere 1997) 
are analyzed through an IDL (RSI) code, to generate images 
to be directly confronted with various kinds of observations.
Rotations, projections, line imaging, diagnostic line ratios maps,
velocity line profiles and PV-diagrams can be performed.

\vspace{-0.3cm}
\section{The luminosity equatorial enhancement}

\begin{figure}
\plottwo{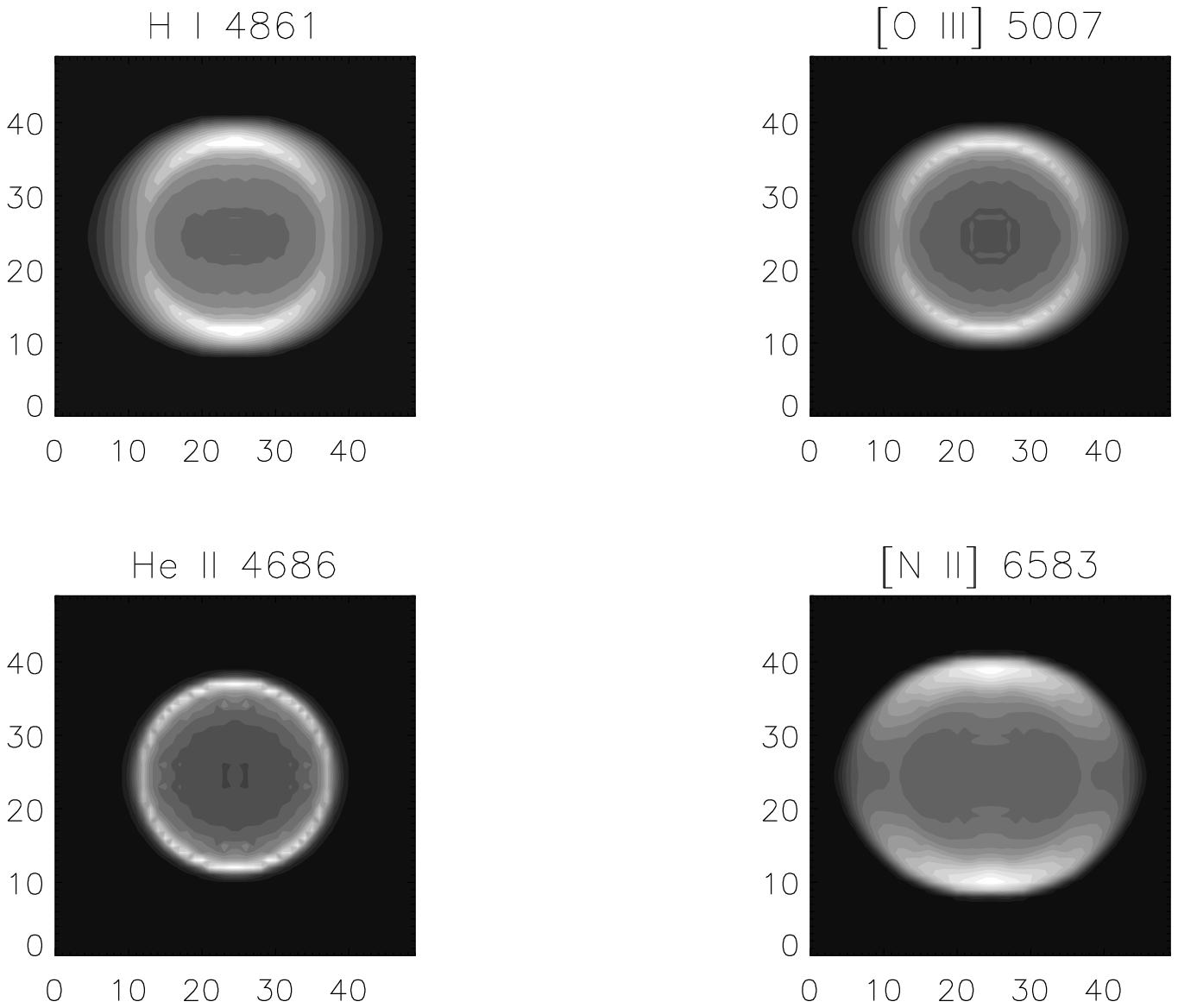}{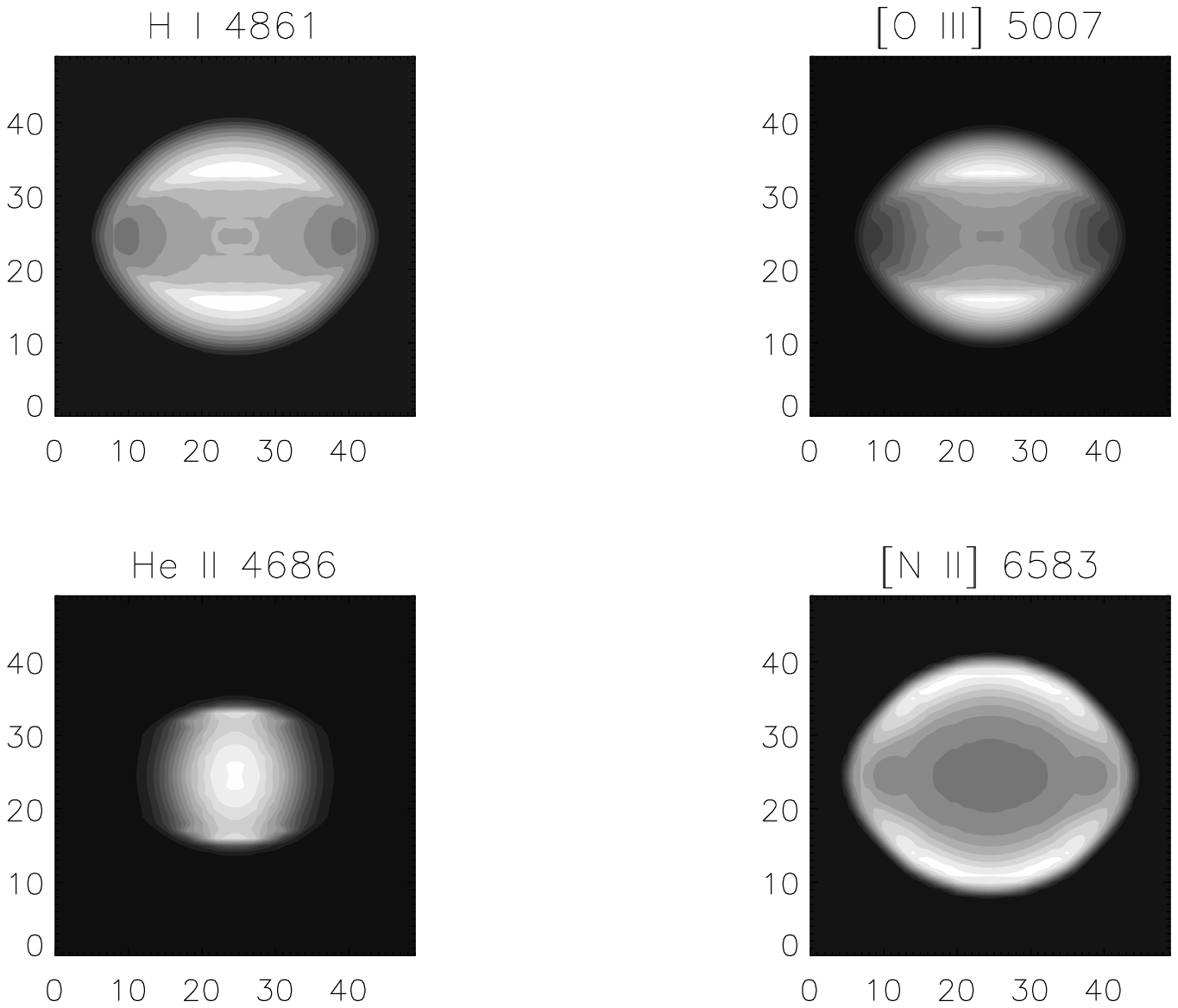}
\caption{Emission-line imaging from Model 1 (four left panels) and 
Model 2 (four right panels). For each model, H$\beta$, [O~III]$\lambda$5007,
He~II$\lambda$4686, and [N~II]$\lambda$6583 intensity maps are shown. 
\label{fig-pap1}
}
\end{figure}

We computed two models of PNe with the same ionizing spectra and the
same mean density, the only difference being the matter distribution:
the first one (Model 1) have a spherical cavity surrounded by a shell with
a density gradient (equatorial enhancement) while the second one (Model
2) is a constant density ellipsoidal shell.  
The global spectroscopic results are similar. The theoretical line images are
displayed in Fig. \ref{fig-pap1}, showing an equatorial luminosity enhancement. 
One way to distinguish
between the two geometries is a density map determined e.g. by the
[S~II]6718/6732 ratio image. See Morisset et al. (2000) for more details.

\vspace{-0.6cm}
\section{A 3D model for NGC 3132}

\begin{figure}
\plottwo{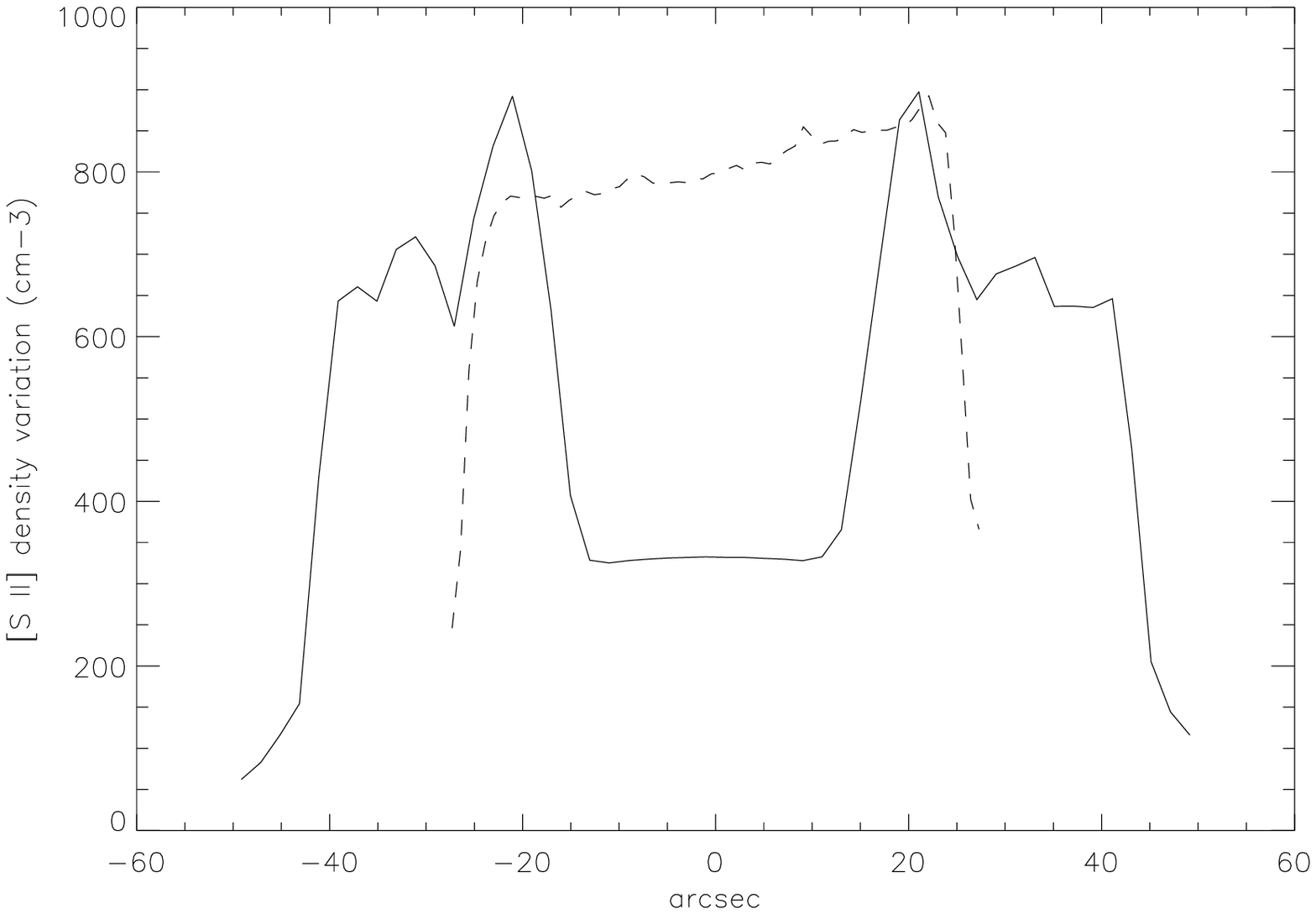}{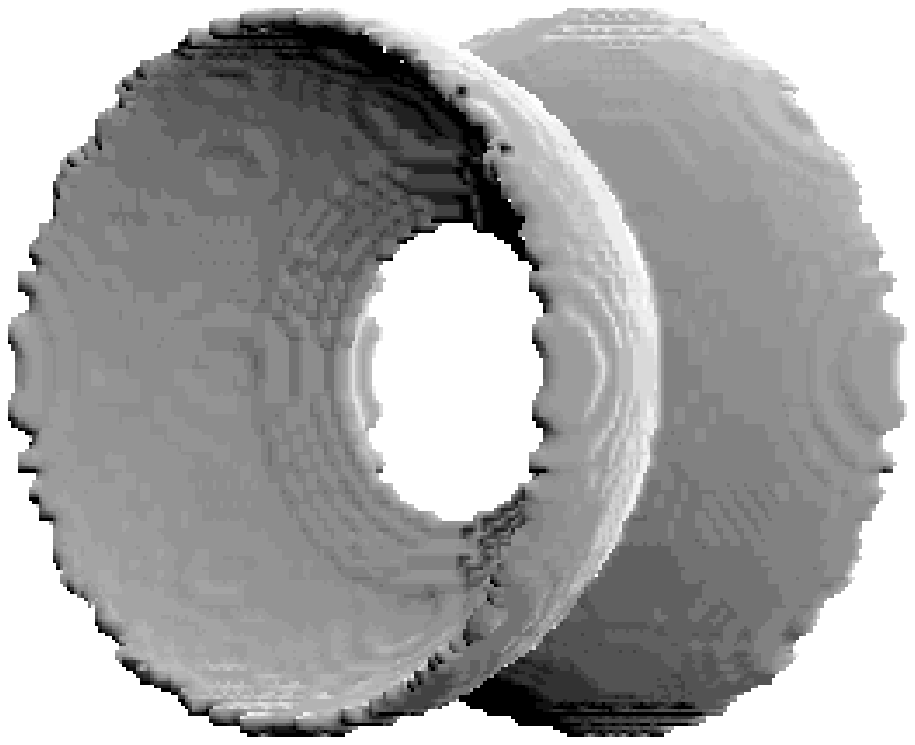}
\caption{
{\bf Left side}: Variation of the electronic density along the
PN, for the Diabolo model (solid line) and the
ellipsoidal model (dashed line).
{\bf Right side}: Diabolo shape. Only high density gas (n$_{\mathrm{H}}$
= 1300 cm$^{-3}$) is shown. The background density is 300 cm$^{-3}$.
\label{fig-diab1}
}
\end{figure}

\begin{figure}
\plotone{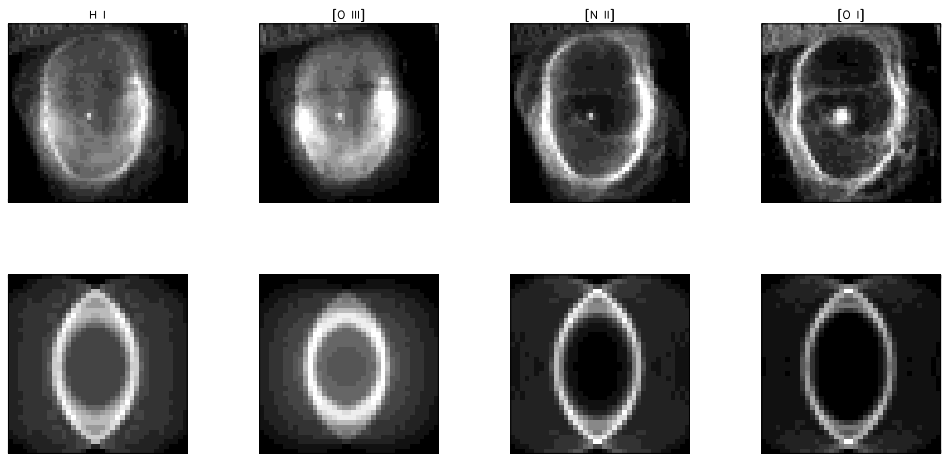}
\caption{HST Observations (top) and corresponding line images from the Diabolo
model (bottom).
\label{fig-diab-hst}
}
\end{figure}

\begin{figure}
\plotfiddle{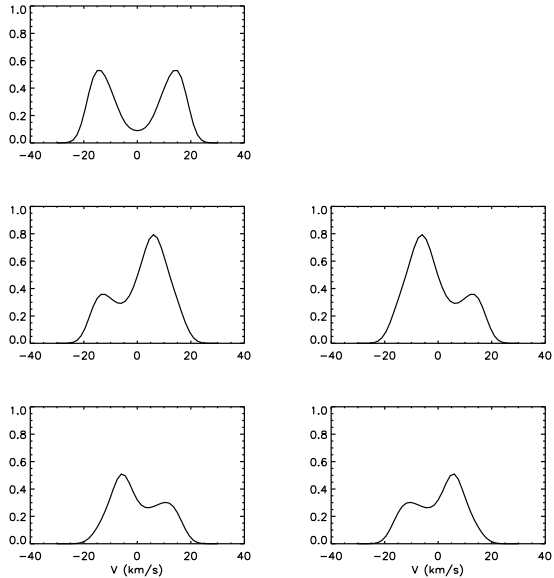}{8.5cm}{0.}{160.}{160.}{-200.}{0.}
\caption{[O~III] Velocity profiles from our Diabolo model, obtained at
the same positions as observed by Sahu \& Desai (1986).
\label{fig-diab_velo}
}
\end{figure}

We applied 3D code to model NGC~3132, classified as an elliptical
PN. We adopt the central star and gas properties used by 
B\"assgen, Diesch \& Grewing (1990) and perform two models differing
in the gas distribution.
We show that an ellipsoidal shell sucessfully
reproduces the low resolution observations (line images and
the global spectroscopy). However, high (spatial and spectroscopic)
resolution observations are not reproduced. 
Observations of [S~II]6718/6732 line ratio indicate a decrease of the eletronic
density towards the center of the PN (Juguet et al. 1988).
The density variation modeled
with the ellipsoidal shell (Fig.2, dashed line) can not reproduce the observed density decrease.
Choosing a Diabolo morphology (Fig. \ref{fig-diab1}, 
right side) we successfully reproduced the central density hole (solid
line, left side).
In Fig. \ref{fig-diab-hst} the line images of NGC~3132 from HST
observations (top four panels) are compared to the results of our
Diabolo model (bottom four panels). The ellipsoidal image
as well as the ionization structure are well reproduced with a 40\deg rotation
of the symmetry axes relative to the line of sight.
Adopting a linear velocity law, we can model the emission line
profiles observed through various apertures. The observed [O~III] profiles
(Sahu \& Desai 1986) can be compared to the theoretical ones obtained with the Diabolo model using their
aperture size and positions ( Fig. \ref{fig-diab_velo}). 
Notice that the asymmetrical profiles observed at the
external part of the PN (four bottom panels) are not obtained with the
ellipsoidal shell model. See Monteiro et al. (2000) for more details

\vspace{-0.4cm}

\vspace{-0.3cm}
\section{Conclusions}

We show that the lack of high resolution observations and
of a 3D modeling tool, can lead erroneous conclusions about the
morphology of PNe. In particular, we showed that NGC~3132 was
misclassified as an elliptical PN. A Diabolo shape is successful in
reproducing: 1) the density decreasing in the central direction of the
PN, 2) the asymmetric [O~III] velocity profiles observed at the
external part of the PN. Both observations are not reproduced by
an ellipsoidal shell model.


\begin{references}
\reference Balick, B. 1987, \aj, 94, 671
\reference B\"assgen, M., Diesch, 
	C. \& Grewing, M. 1990, BDG90, A\&A 237, 201 
\reference Corradi, R. L. M. \& 
	Schwarz, H. E. 1995, \aap, 293, 871
\reference Gruenwald, R., Viegas, S. M. \& Brogui\`ere, D. 1997, 
	ApJ, 480, 283
\reference Juguet, J. L., Louise, R., Macron, A. \& Pascoli, G. 1988,
	A\&A, 205, 267 
\reference Kwok, S., Purton, C. R. \& Fitzgerald, P. M. 1978, \apj 219,
	L125
\reference Monteiro, H., Morisset, C., Gruenwald, R., \& Viegas,
	S.M. 2000, \apj, submitted
\reference Morisset C., Gruenwald, R. \& Viegas, S.M. 2000, \apj, submitted
\reference Sahu, K. C. \& Desai, J. N. 1986, A\&A, 161, 357
\reference Stanghellini, L., Corradi,
	R. L. M., \& Schwarz, H. E. 1993, \aap, 279, 521
\end{references}
\end{document}